\documentclass[a4paper,10pt, twoside]{article}
\pdfoutput=1
\usepackage{authblk}
\usepackage{graphicx}
\usepackage{amsmath} 
\usepackage{ae} 
\usepackage{aecompl} 
\usepackage{amssymb}
\usepackage[english]{babel} 
\usepackage{color}
\usepackage{cite}
\usepackage{verbatim}
\usepackage{placeins}

\title{ {\bf Surface properties of liquid mercury: a comparison of density-dependent and density-independent force fields}}
\author[1]{Anton Iakovlev\footnote{Email address: \texttt{iakovlev@theorie.physik.uni-goettingen.de}}}
\author[2]{Dmitry Bedrov}
\author[1]{Marcus M\"{u}ller}
\affil[1]{{\small \emph{ Institut f\"{u}r Theoretische Physik, Georg-August-Universit\"{a}t G\"{o}ttingen, Friedrich-Hund-Platz 1, 37077 G{\"o}ttingen, Germany}}}
\affil[2]{{\small \emph{ Department of Materials Science \& Engineering, University of Utah, 122 South Central Campus Dr., Salt Lake City, 84112, Utah, USA}}}

\date{March 14, 2015}

\begin{document}



\maketitle

\begin{abstract}
Motivated by an experimental interest we investigate by the means of atomistic Molecular Dynamics simulation the ability of density-independent, empiric density-dependent, and recently proposed embedded-atom force fields
for liquid mercury to predict the surface tension of the free surface of liquid mercury at the temperature of 293~K. The effect of the density dependence of the studied models on the liquid-vapor coexistence and surface tension is 
discussed in detail. In view of computational efficiency of the density-independent model we optimize its functional form to obtain higher surface tension values in order to improve agreement with experiment. 
The results are also corroborated by Monte Carlo simulations and semi-analytic estimations of the liquid-vapor coexistence density.
\end{abstract}

\section{Introduction}
\label{sec:intro}
Mercury is a ubiquitous element, which is used as a main constituent and/or produced as a byproduct of major technical processes crucial for the contemporary functioning of human civilization. 
Some of the numerous areas, which involve mercury (Hg) are energy production\cite{saarnio}, gold mining \cite{toxic}, and fluorescent bulbs \cite{lamp}.
Main difficulties related to the usage of mercury are the reduction of Hg-containing residues release into the environment and the elimination of Hg-contamination \cite{saarnio, toxic, selin}. 
Mercury can be readily bound by incorporating it within plate-like crystals of Hg(S-R)$_{2}$ \cite{pokroy}.
Several other organic molecular systems have also been recently designed, which are capable of efficiently localizing mercury for subsequent disposal \cite{tan, tao, disposal}. 
On the other hand, liquid mercury represents a particular interest in electrowetting applications enabling the construction of highly conducting metallic nanowires inside carbone nanotubes, 
which is actively studied by both experiment and computer simulation \cite{zhao, kutana1, kutana2, kutana3}.
Because of its seamless surface liquid mercury also enables the creation of high quality defectless self-assembled monolayers (SAMs) of various organic molecules \cite{babayco, love, demoz}. 
Therefore the surface of liquid mercury serves as a model system for the experimental study of universal (free of the underlying crystal substrate) properties of numerous organic films 
\cite{kraack1, ocko1, deutsch1, magnussen1, kraack2, kraack3, kraack4, kraack5, kraack6, kraack7, stevenson}.
Moreover, the unique properties of liquid mercury, especially its high surface tension, make Hg droplets superior to other metals for producing metal-SAM-metal \cite{babayco, holmlin} 
and metal-SAM-semiconductor \cite{zhu, popoff} junctions. This property motivates the wide usage of liquid mercury in organic electronics \cite{holmlin, zhu, popoff, weiss, rampi, nitzan, slowinski1, slowinski2}, 
and even in the design of mixed organic-metal systems for information processing \cite{kiehl}. 
Electric properties of Hg-SAM junctions are typically tunable by the shape of a sessile Hg droplet or one that is suspended from a pipette and the surface coverage of SAMs (see e.g. \cite{zhu}).
A better control of the contact angle and of the surface tension of Hg droplets, respectively, would enable a refined control over the shape, 
and thus over other relevant properties of liquid Hg in pipettes, carbon nanotubes and/or on a substrate. 

In practice, however, the design of appropriate systems involving Hg compounds is extremely dangerous and arduous because of the mercury's high toxicity \cite{toxic, clarkson, zahir}.
Therefore the ability to model and predict the properties of hybrid surfactant-mercury systems as well as pure mercury surfaces by computer simulation may substantially 
facilitate further progress in the above mentioned areas. 
Coarse-grained MD computer simulation techniques, where the quantum degrees of freedom have been integrated out, offer a tractable strategy to study the statistical mechanics of SAM systems on crystalline substrates
being capable of explaining a number of experimental observations \cite{filippini, ahn, jimenez, ghorai, singh, henz, rai, vemparala, zhang, luedtke, bhatia1, bhatia2, tupper, sellers, hautman}.
In this context, an effective force field compatible with the MD framework and yielding a satisfactory thermodynamic description of the free liquid mercury surface (i.e. high surface tension and  dense liquid) at experimentally important temperatures 
is the necessary starting point for the large-scale MD simulation of SAM-Hg
or other interfacial systems and phase coexistences involving liquid mercury.
The need to reproduce unique properties of liquid Hg has generated over the last decades a number of works \cite{desgranges, belashch1, bomont1, belashch2, bomont2, raabe1, ghatee, raabe2, toth, okumura, belashch3, chacon, sumi, munejiri}
all underlining the importance of the state-dependent interactions, either through temperature or density, to mimic intricate many-body effects due to the complicated underlying electron structure of liquid Hg. 
As was previously discussed by Louis, an empirical inclusion of a density dependence in a pair potential requires additional care when used \cite{louis}.
In this article we analyze the ability of relatively simple density-independent (DI) \cite{chacon}, empirical density-dependent (DD) \cite{bomont1}, and recently proposed embedded-atom (EAM) \cite{belashch1, belashch2} interaction models to
predict the surface tension of the free surface of liquid Hg at $T = 293$~K. In the next section we start by describing the interaction models and our computational protocol, then we proceed with the discussion of obtained results, and conclude
with some final remarks.

\section{Methods and models}
\label{sec:methods}
\subsection{Pair force fields}
\subsubsection{Density-Independent Model}
\label{sec:didd}
A naive guess would be to use a Lennard-Jones (LJ) model $\Phi_{LJ}(r)=4\epsilon[(\sigma/r)^{12} - (\sigma/r)^{6}]$ ($\sigma$ and $\epsilon$ being the characteristic length and energy 
scales of the LJ potential) to model the liquid mercury at low temperatures, but it would yield a too low surface tension. 
Let us, for example, consider the LJ system at the triple point, since at this thermodynamic state a liquid would have the highest surface tension $\gamma$.
For the LJ liquid near to the triple point holds $\gamma^* = \gamma \sigma^2/\epsilon = 1.137$ \cite{shen}, $T^* = k_{\rm{b}} T/\epsilon = 0.7$ and 
$\rho*=\rho \sigma^3 = 0.8$ \cite{nijmeijer, johnson}. 
Assuming the LJ system has the same triple point as the liquid Hg at $T = 235$~K and $\rho = 13.69$~g/cm$^3$ one can recover $\epsilon$ and $\sigma$, 
and obtain the LJ prediction $\gamma \approx 0.07$~N/m, which is considerably smaller than even the experimental value of $\sim$0.5~N/m at $T = 293$~K \cite{gast}. 
This confirms the inappropriateness of the LJ-like potentials for simulating the liquid mercury at low temperatures close to the triple point \cite{kutana4, hoshino}

The DI pair potential for liquid mercury was given by \cite{chacon}
\begin{equation} 
 \Phi(r) = A_0 e^{-a r} - A_1 e^{-b(r-R_0)^2}.
 \label{eq:DI}
\end{equation}
The values of the parameters at temperature $T=293$~K are $A_0 = 8.2464\times10^{13}$~eV, $a = 12.48$~\r{A}$^{-1}$, $b = 0.44$~\r{A}$^{-2}$, $R_0 = 3.56$~\r{A} 
and $A_1 = 0.0421$~eV \cite{bomont2}. The first term describes the Born-Meyer repulsion. The second term in Eq.~\ref{eq:DI} determines the attractive potential with a well depth $A_1$, which is in general temperature-dependent,
and minimum at $R_0$. The parameter $b$ is responsible for the range of the attractive interaction.

\subsubsection{Density-Dependent Model}
The empirical DD pair potential can also be defined by Eq.~\ref{eq:DI} but, additionally, in order to account for the metal-nonmetal transition with the change of density
the coefficient $A_1$ is made density-dependent in the following way \cite{bomont1}
\begin{equation}
 A_1 = f_i f_j A_1^{MM} + \left[f_i \left(1 - f_j\right) + \left(1 - f_i\right) f_j \right] A_1^{MV}
      + \left(1 - f_i\right) \left(1 - f_j\right) A_1^{VV}, \label{eq:DD}
\end{equation}
where $A_1^{MM} = 0.0421$~eV, $A_1^{MV} = 0.0842$~eV, $f_i \equiv f(\rho(\mathbf{r_i}))$ is a function of the local density $\rho(\mathbf{r_i})$ at the position of \textit{i}-th atom, and is given by 
$f(\rho) = 1$ if the density is larger than a threshold value 11~g/cm$^3$ and it vanishes for $\rho<8$~g/cm$^3$, and is chosen to be a smooth function in between, i.e., only if the local density around a particle is in 
the intermediate density interval between 8 and 11~g/cm$^3$ does the density dependence of the potential matter.  
We have tested polynomials of the 1st, 2nd, 3rd and 5th order to represent $f(\rho)$, and found that the functional form of $f(\rho)$ is irrelevant for the phase behavior. In the following we use the 3rd
order polynomial for $f(\rho)$.

The way to compute the local density $\rho(\mathbf{r})$ must not depent on geometry to enable the simulations of experimentally relevant systems, for instance systems with the varying shape of Hg droplets (see e.g. Ref~\cite{zhu}).
Thus we adopt standard weighting functions $w$ to estimate $\rho(\mathbf{r})$ \cite{rapaport}
\begin{equation}
 \rho(\mathbf{r_i}) = \sum_{j \ne i} w(r_{ij}),
 \label{eq:locdens}
\end{equation}
with $r_{ij}=\sqrt{(\mathbf{r_i}-\mathbf{r_j})^2}$ being the relative distance between atoms $i$ and $j$, and normalization
\begin{equation}
 4\pi \int_0^{r_{\rm{c}}} dr \, r^2 w(r) = 1,
 \label{eq:wnorm}
\end{equation}
where $r \equiv r_{ij}$ and $r_{\rm{c}} = 9$~\r{A} is the cut-off radius of the pair potential. 
We have carried out simulations with different values of $A_1^{VV}$ and we have chosen $A_1^{MM}=A_1^{VV}$ in order to obtain consistent results in the limit of low and high densities.

\subsection{EAM force fields}

The Embedded-Atom Method (EAM) represents a general framework of accounting for many-body effects of the underlying electron structure of metals in atomistic simulations \cite{foiles1, foiles2, holzman}. 
A potential energy per atom \textit{i} at the position $\mathbf{r_i}$ is given by
\begin{equation}
 \Phi_{i} = \Phi_{\rm{em}}(\rho_{\rm{dl}}(\mathbf{r_i})) + \sum_{j \ne i}\Phi(r_{ij}),
 \label{eq:phi_i_eam}
\end{equation}
where $\Phi_{\rm{em}}$ is the embedded energy due to the many-body electronic interactions, $\rho_{\rm{dl}}$ is an effective electron density at position $\mathbf{r_i}$, and $\Phi(r)$ is the pair interaction between the ions of a metal.
Currently there exist two parameterizations of the EAM model for the liquid mercury \cite{belashch1, belashch2}, which essentially differ in the form of the embedded energy $\Phi_{\rm{em}}$. For the first one the embedding energy 
is given by \cite{belashch1}
\begin{subequations}
 \begin{align}
  \Phi_{\rm{em}} =& \, a_1 + c_1(\rho_{\rm{dl}} - \rho_0)^2, \; \rho_1 < \rho_{\rm{dl}} \le \rho_8, \label{eq: eam1} \\
  \Phi_{\rm{em}} =& \, a_i + b_i(\rho_{\rm{dl}} - \rho_{i-1}) + c_i(\rho_{\rm{dl}} - \rho_{i-1})^2, \; \rho_i < \rho_{\rm{dl}} \le \rho_{i-1}, \; i = 2, \, \dots, \, 7, \label{eq:eam2} \\
  \Phi_{\rm{em}} =& \, \left[ a_8 + b_8(\rho_{\rm{dl}} - \rho_7) +c_8(\rho_{\rm{dl}} - \rho_7)^2 \right] \left[ 2 \frac{\rho_{\rm{dl}}}{\rho_7} - \left( \frac{\rho_{\rm{dl}}}{\rho_7} \right)^2 \right], \; \rho_{\rm{dl}} \le \rho_7, \label{eq:eam3} \\
  \Phi_{\rm{em}} =& \, a_9 + b_9(\rho_{\rm{dl}} - \rho_8) + c_9(\rho_{\rm{dl}}-\rho_8)^m, \; \rho_8 < \rho_{\rm{dl}} \le \rho_9, \label{eq:eam4} \\
  \Phi_{\rm{em}} =& \, a_{10} +b_{10}(\rho_{\rm{dl}} - \rho_9) + c_{10}(\rho_{\rm{dl}} - \rho_9)^n, \; \rho_{\rm{dl}} > \rho_9. \label{eq:eam5}
 \end{align}
\end{subequations}
And for the second one $\Phi_{\rm{em}}$ takes the form \cite{belashch2}
\begin{subequations}
 \begin{align}
  \Phi_{\rm{em}} =& \, a_1 + a_2(\rho_{\rm{dl}}-\rho_0)^2 + a_3(\rho_{\rm{dl}}-\rho_0)^3, \; \rho_{\rm{dl}} \ge 0.8\rho_0, \label{eq:eam6} \\
  \Phi_{\rm{em}} =& \,  a_4\sqrt{\rho_{\rm{dl}}} + a_5\rho_{\rm{dl}}, \; \rho_{\rm{dl}} < 0.8\rho_0. \label{eq:eam7}
 \end{align}
\end{subequations}
The effective density $\rho_{\rm{dl}}$ is determined by
\begin{equation}
 \rho_{\rm{dl}}(\mathbf{r_i}) = \sum_{j \ne i} \psi(r_{ij}),
 \label{rhodl}
\end{equation}
with
\begin{equation}
 \psi(r) = p_1 \exp \left[ -p_2 r \right].
 \label{eq:psieam}
\end{equation}
In the following we shall denote the first and the second parameterization models as EAM2013 and EAM2006, respectively.
\begin{figure}[h!]
	\centering
	\includegraphics[width=100mm]{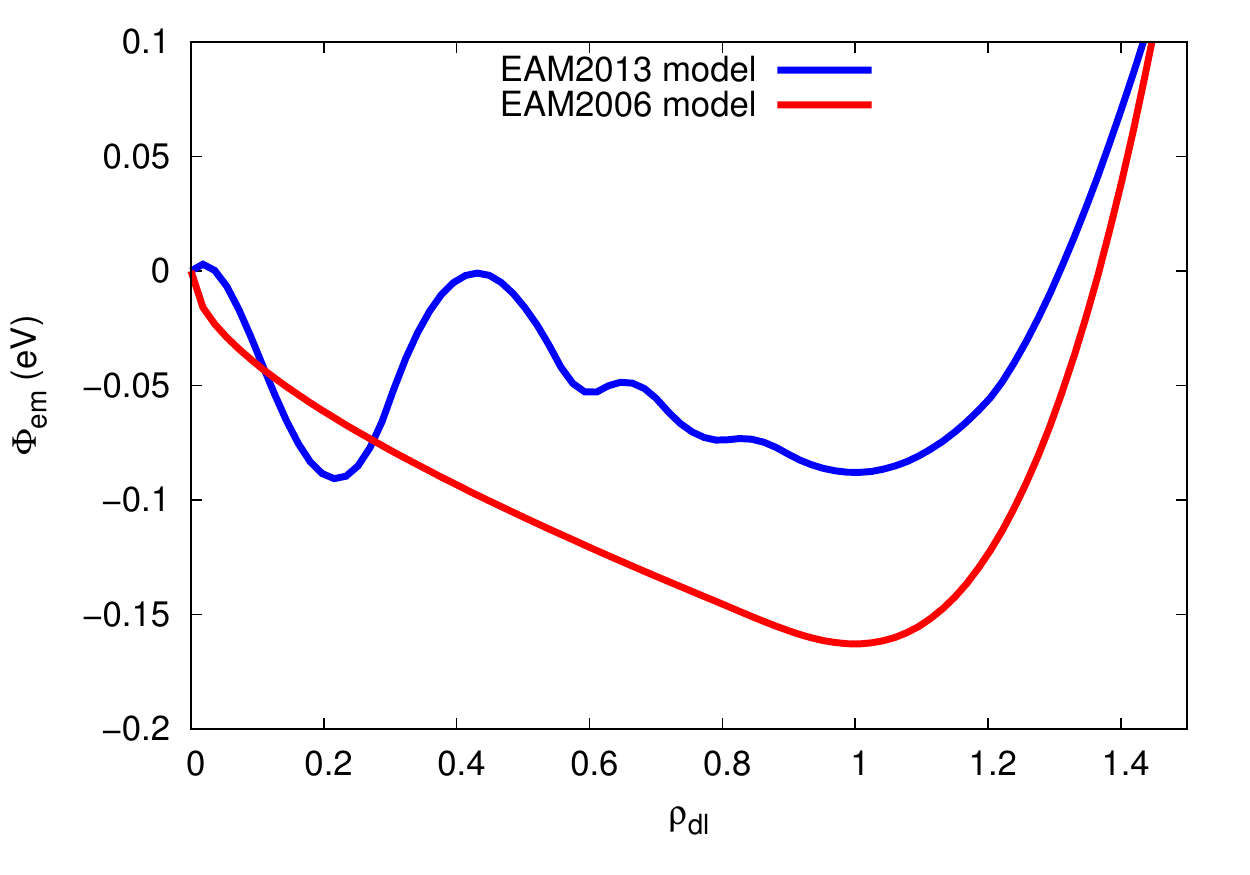}
	\caption{The embedding energy for EAM2013 and 2006 models.}
      	\label{fig:phi_of_rho_eam}
\end{figure}
The embedding energy for the EAM2013 and EAM2006 models is shown in Fig.~\ref{fig:phi_of_rho_eam}. The EAM2013 model is specifically designed for the case of the strong compression of liquid mercury.
The meaning and values of the rest of the parameters for these two models the reader can find in Refs.~\cite{belashch1, belashch2}, respectively.
The effective pair potential $\Phi(r)$ at $T = 293$~K for both models for $ 2.55 \le r \le 8.35$~\r{A} is tabulated in Ref.~\cite{belashch2}, and
for $0 \le r < 2.55$~\r{A} is defined as \cite{belashch1}
\begin{equation}
 \Phi(r) = \alpha_1 - \alpha_2 (2.55 - r) + \alpha_3 \left[ \exp(\alpha_4 (2.55 - r)) - 1 \right].
\end{equation}
The values for the parameters $\alpha_1$, $\alpha_2$, $\alpha_3$ and $\alpha_4$ are given in Ref.~\cite{belashch1}.
At the cut-off radius $r_{\rm{c}} = 8.35$~\r{A} the pair potential is set to zero. The minimum of $\Phi(r)$ is -0.0617~eV at $r_{\rm{min}} = 3.18$~\r{A}.

\subsection{Molecular Dynamics Simulation}
\label{sec:md}
The LAMMPS package \cite{lammps} with model specific extensions is used to carry out MD simulations of the mercury film. 
Surface configurations are prepared by the following procedure. 
For the DI and  DD models we first equilibrated the bulk Hg for 8 ns. For the EAM2013 and EAM2006 models the initial bulk configurations were equilibrated at $T = 293$ K for 48 ns.

The sizes of the computational cell parallel to the film's surface are  set equal to 24.56~\r{A} for all the models, 
and in the perpendicular direction to 80.673 \r{A} (89.0595 \r{A}) for the DI and DD (EAM2013 and EAM2006) models.
Periodic  boundary conditions are applied in all direction. 
Second, we add empty space above and below the Hg film along the z-axis, so that the film would not feel the presence of walls if they were placed at both ends of the system along the z-axis. 
The film's center of mass is placed in the middle of the slab. 
Afterwards the film is equilibrated for 2 ns at $T = 293$~K with periodic boundary conditions applied in all directions and the measurements are taken every 10 fs during the total simulation time of 16 ns for all the models. 
The time step of 1~fs and the total number of Hg atoms $N = 2000$ are used through out this paper. 
Since, in the current study, we are not interested in dynamics, a simple temperature rescaling is used to control the temperature. 
A trial simulation with Nos\'{e}-Hoover thermostat \cite{nose, hoover, chains} yielded virtually identical results.

The calculation of forces for the DI pair potential (Eq.~\ref{eq:DI}) is straightforward. For the DD case one can show that the force between two particles can also be effectively represented as a pair force, 
which is very useful for the practical implementation. To this end, we start with the definition of the total force $\mathbf{F_k}$ acting on \textit{k}-th particle at the position given by a radius vector $\mathbf{r_k}$
\begin{equation}
 \mathbf{F_k} = -\mathbf{\nabla_k}U,
 \label{eq:force_def}
\end{equation}
where $U$ is the total potential energy of the system
\begin{equation}
 U = \frac{1}{2}\sum_{i,j\ne i} \Phi_{ij},
 \label{eq:tot_e}
\end{equation}
with $\Phi_{ij}\equiv\Phi(r_{ij})$.
For the x-component $F_{k_x}$ of $\mathbf{F_k}$ we obtain
\begin{equation}
 F_{k_x} = - \frac{\partial U}{\partial x_k} = F_{k_x}^{di} + F_{k_x}^{dd},
 \label{eq:force_x}
\end{equation}
where $F_{k_x}^{di}$ and $F_{k_x}^{dd}$ are given by 
\begin{subequations}
\begin{align}
 F_{k_x}^{di} &= -\frac{1}{2} \left[ \sum_{i,j\ne i} \frac{\partial\Phi_{ij}}{\partial x_i}\delta_{ik} + \sum_{i,j\ne i}\frac{\partial\Phi_{ij}}{\partial x_j}\delta_{kj} \right], \label{eq:fx_di} \\
 F_{k_x}^{dd} &= \frac{1}{2}\sum_{i,j\ne i} \frac{\partial A_1}{\partial x_k}\varphi_{ij}, \label{eq:fx_dd}
\end{align}
\end{subequations}
where we introduced the notation $\varphi_{ij}\equiv\exp[-b(r_{ij}-R_0)^2]$ and $\delta_{ik}$ denotes the Kronecker delta.

Eq.~\ref{eq:fx_di} yields the x-component of the force on the \textit{k}-th particle exactly for the DI case, i.e., when the coefficient $A_1$ is density-independent.
The calculation of $F_{k_x}^{di}$ is straightforward. Using the Kronecker deltas we first eliminate the summation over indices \textit{i} and \textit{j} in Eq.~\ref{eq:fx_di} in the 1st and 2nd sums respectively
\begin{equation}
 F_{k_x}^{di} = -\frac{1}{2} \left[ \sum_{j\ne k} \frac{\partial\Phi_{kj}}{\partial x_k} + \sum_{i\ne k}\frac{\partial\Phi_{ik}}{\partial x_k} \right] = -\sum_{i\ne k} \frac{\partial\Phi_{ki}}{\partial x_k}.
 \label{eq:fx_di1}
\end{equation}
Finally after evaluating the derivative we obtain
\begin{equation}
 F_{k_x}^{di} = \sum_{i\ne k} \left[ \alpha A_0 e^{-\alpha r_{ki}} - 2\beta(r_{ki} - R_0)
                A_1e^{-b(r_{ki}-R_0)^2} \right] \frac{\Delta x_{ki}}{r_{ki}},
 \label{eq:fx_di2}
\end{equation}
where $\Delta x_{ki}=x_k-x_i$. Now in order to evaluate $F_{k_x}^{dd}$ we have to consider the many-body contribution stemming from the density-dependence of $A_1$ on the local density $\rho(\mathbf{r})$ at the 
positions of the two respective particles that interact. By differentiating Eq.~\ref{eq:DD} w.r.t. $x_k$ and inserting the result of the differentiation into Eq.~\ref{eq:fx_dd} we obtain
\begin{equation}
 F_{k_x}^{dd} = \sum_{i,j\ne i} \varphi_{ij}\frac{df_i}{d\rho_i}\frac{\partial\rho_i}{\partial x_k} (f_jB + C),
 \label{eq:fx_dd1}
\end{equation}
where
\begin{subequations}
\begin{align}
  B &= A_1^{MM} - 2A_1^{MV} + A_1^{VV}, \label{eq:B} \\
  C &= A_1^{MV} - A_1^{VV}. \label{eq:C}
\end{align}
\end{subequations}
Giving Eq.~\ref{eq:locdens} we have
\begin{subequations}
 \begin{align}
  F_{k_x}^{dd} =& \sum_{i,j\ne i} \varphi_{ij}\frac{df_i}{d\rho_i}(f_jB + C)
	          \sum_{l\ne i}\left[ \frac{dw_{kl}}{dr_{kl}}\frac{dr_{kl}}{dx_k}\delta_{ki} + \frac{dw_{ik}}{dr_{ik}}\frac{dr_{ik}}{dx_k}\delta_{kl}\right] \label{subeq:fx_dd2_1}\\
               =& \sum_{j\ne k} \varphi_{kj} \frac{df_k}{d\rho_k}(f_jB + C) \sum_{l\ne k} \frac{dw_{kl}}{dr_{kl}}  \frac{\Delta x_{kl}}{r_{kl}} \nonumber \\
	        & + \sum_{i\ne k, j\ne i} \varphi_{ij} \frac{df_i}{d\rho_i}(f_jB + C)\frac{dw_{ik}}{dr_{ik}}\frac{-\Delta x_{ik}}{r_{ik}}.  
 \end{align}
 \label{subeq:fx_dd2_2}
\end{subequations}
By relabeling the index \textit{l} with \textit{i} we obtain
\begin{equation}
 F_{k_x}^{dd} = \sum_{i\ne k} y_{ki}\frac{dw_{ki}}{dr_{ki}}\frac{\Delta x_{ki}}{r_{ki}},
 \label{eq:fx_dd3}
\end{equation}
with a shorthand notation
\begin{equation}
 y_{ki} = \frac{df_k}{d\rho_k}\sum_{j\ne k}\varphi_{kj}(f_jB + C) + \frac{df_i}{d\rho_i} \sum_{j\ne i}\varphi_{ij}(f_jB + C).
 \label{eq:yik}
\end{equation}
We note that $y_{ki}=y_{ik}$, which basically enables us finally to represent the x-component (Eq.~\ref{eq:force_x}) of the total force on the \textit{k}-th particle as a sum of effective pair forces that obey the Newton's 3rd law
\begin{equation}
 F_{k_x} = \sum_{i\ne k} f_{ki},
 \label{eq:f_tot_pair}
\end{equation}
where $f_{ki}$ is the effective pair force that the \textit{i}-th Hg atom exerts on the \textit{k}-th one. Taking into account Eqs.~\ref{eq:force_x},~\ref{eq:fx_di2}~and~\ref{eq:fx_dd3}, we write $f_{ki}$ in the form
\begin{equation}
 f_{ki} = \left[ \alpha A_0 e^{-\alpha r_{ki}} - 2b(r_{ki}-R_0)A_1 e^{-b(r_{ki}-R_0)^2}
         + y_{ki}\frac{dw_{ki}}{dr_{ki}} \right] \frac{\Delta x_{ki}}{r_{ki}}.
 \label{eq:f_pair}
\end{equation}
Similar expressions hold for the y- and z-components of the total force $\mathbf{F_k}$ (Eq.~\ref{eq:force_def}).

The calculation of forces for the EAM2013 and EAM2006 models are given in Ref.~\cite{belashch2}.

The surface tension $\gamma$ is computed in the MD simulation by using the Kirkwood-Buff relation \cite{kirkwood}
\begin{equation}
 \gamma = \frac{1}{2} \int dz \left( p_n - p_{\tau} \right),
 \label{eq:gamma}
\end{equation}
where $p_n = p_{zz}$ and $p_{\tau} = (p_{xx} + p_{yy})/2$ are normal and tangential w.r.t.\ the liquid Hg film's surface stresses given by the diagonal elements of the stress tensor $p_{xx}$, $p_{yy}$ and $p_{zz}$, 
which are the pressure components in x-, y- and z-direction respectively (see e.g.\ \cite{tildesley}). For the \textit{on-the-fly} numeric calculation of $\gamma$ we have used the discretization scheme of Eq.~\ref{eq:gamma} 
described in Ref.~\cite{nijmeijer}. Giving the large cut-off radii used in all the treated models we assume the tail corrections to $\gamma$ to be small.

\subsection{Monte Carlo simulation}
To validate the MD simulations with the DI and DD models we have carried out MC simulations using the Metropolis algorithm \cite{metropolis} with a Verlet neighbor list \cite{tildesley}. The maximal local displacement for each atom 
along any axis is 0.15~\r{A}  so that the acceptance ratio is approx.\ 56\% and 73\% for the DI and DD models, respectively. The initial Hg film is equilibrated for $3\times 10^4$ MC sweeps, where each of $2\times10^3$ atoms was attempted to move once in 1 MC sweep. 
Then density profiles are averaged over $1.045\times10^6$ and $5.35\times10^5$ sweeps for the DI and DD force fields, respectively. 

\subsection{Liquid State Theory}
\label{sec:LST}
The MD simulation results are complemented by semi-analytic thermodynamic considerations. 
Similar approaches have proven useful to predict the equation of state of liquid mercury \cite{kitamura}.
It is based on the Barker-Henderson  perturbation theory to calculate the Helmholtz free energy \cite{barker}
\begin{equation}
 \frac{\beta F}{V} = \rho \ln \frac{\rho \Lambda_{\rm{T}}^3}{e} + \rho \frac{4\eta - 3\eta^2}{(1 - \eta)^2}
                     + \frac{\rho^2}{2} \int_0^{r_{\rm{c}}} d^3r \beta \Phi_{a}(r) g_{\rm hs}(r) + F_{\rm{em}}, 
 \label{eq:free_en}
\end{equation}
where $V$ is the system's total volume, $\beta = 1/(k_{\rm{b}}T)$, $k_{\rm{b}}$ is the Boltzmann constant, $\Lambda_{\rm{T}} = (2\pi\hbar^2\beta/m)^{\frac{1}{2}}$ is the thermal de-Broglie wavelength,  $m$ denotes the Hg atom mass, 
$\eta=\pi\rho d^3/6$ is a packing fraction, $d$ is an effective hard-sphere (HS) diameter of the Hg atom, $g_{\rm hs}(r)$ is the HS pair correlation function \cite{chang}.
The first terms in Eq.~\ref{eq:free_en} is the ideal gas contribution \cite{landau}. The second term is the free energy of the reference repulsive HS system obtain from the Carnahan-Starling equation of state \cite{carnahan}. 
We also apply the routinely used Weeks-Chandler-Andersen (WCA) decomposition of the pair potential $\Phi(r)$ into the reference repulsive part $\Phi_{\rm{r}}$ and pure attractive interaction $\Phi_{\rm{a}}$ as \cite{WCA_science, weeks}
\begin{equation}
 \Phi(r) = \Phi_{\rm{r}}(r) + \Phi_{\rm{a}}(r),
 \label{eq:wca}
\end{equation}
where
\begin{equation}
 \Phi_{\rm{r}}(r) =
 \begin{cases}
 \Phi(r) - \Phi(r_{\rm{min}}) & \textnormal{if $r \le r_{\rm{min}}$},\\
 0 & \textnormal{otherwise},
 \end{cases}
 \label{eq:phi_r}
\end{equation}
and
\begin{equation}
 \Phi_{\rm{a}}(r) =
 \begin{cases}
  \Phi(r_{\rm{min}}) & \textnormal{if $r \le r_{\rm{min}}$},\\
  \Phi(r) & \textnormal{otherwise},
 \end{cases}
 \label{eq:phi_a}
\end{equation}
with $r_{\rm{min}}$ being the minimum of the pair potential $\Phi(r)$, and the attraction is then treated perturbatively in the third term of Eq.~\ref{eq:free_en}.
The effective HS diameter is given then by \cite{barker}
\begin{equation}
 d = \int_0^{r_{\rm{min}}} dr \left(  1 - e^{-\beta \Phi_{\rm{r}}(r)} \right).
 \label{eq:dhs}
\end{equation}
Finally, the fourth term in Eq.~\ref{eq:free_en} is the embedding energy contribution to the free energy, which is exact within the current formalism and non-zero only for the EAM2013 and EAM2006 models
\begin{equation}
 F_{\rm{em}} = \beta \rho \Phi_{\rm{em}}(\langle \rho_{\rm{dl}} \rangle),
 \label{eq:Fem}
\end{equation}
where the bulk average effective dimensionless density $\langle \rho_{\rm{dl}} \rangle$ is calculated from the particle density $\rho$ as \cite{belashch2}
\begin{equation}
 \langle \rho_{\rm{dl}} \rangle = 4\pi \rho \int_0^{\infty} g(r) \psi(r) r^2 dr,
 \label{eq:rho_to_rhodl}
\end{equation}
with $\rho$ in the units of \r{A}$^{-3}$. In our calculations we approximate $g(r)$ with the one for the hard spheres, which appears to be a very good approximation in this case, 
since in such a way calculated $\langle \rho_{\rm{dl}} \rangle$ deviates only within 0.7\% from the corresponding value obtained by using the experimental $g(r)$ at $T = 293$~K \cite{tamura}.

Giving Eq.~\ref{eq:free_en} one can now estimate the free energy for a desired thermodynamic state characterized by a given pair of $\rho$ and $T$ values.
In order to calculate the coexistence density we construct the free energy vs density curve, $F(\rho)$, and use a double-tangent construction \cite{cahn} as illustrated in the following section.

\section{Results and discussions}
\label{sec:results}
\subsection{Density-Independent Models}
\begin{figure}[h]
	\centering
	\includegraphics[width=100mm]{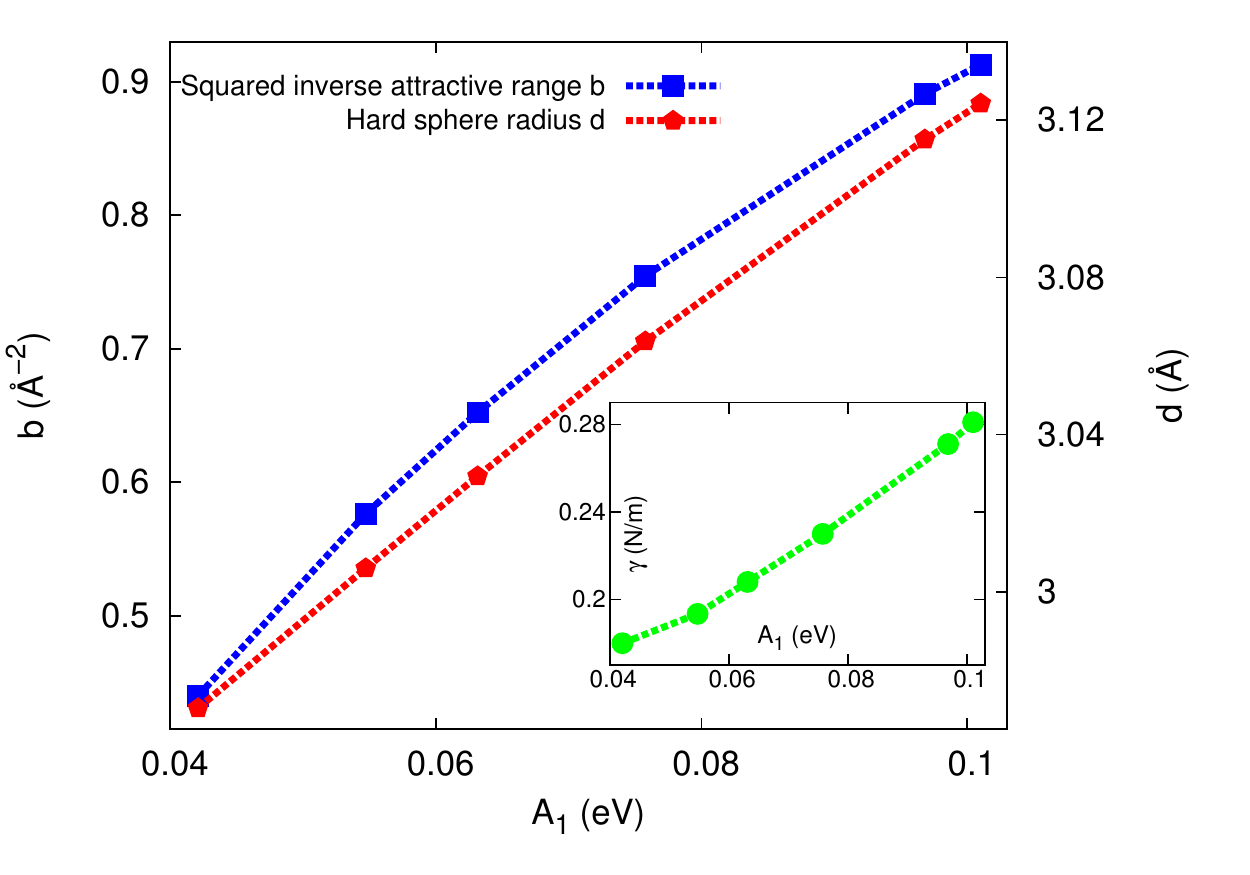}
	\caption{The squared inverse attractive range $b$ (left y-axis, blue squares), the hard core radius $d$ (right y-axis, red pentagons) and the surface tension $\gamma$ 
	(inset, green circles) are shown as functions of the interaction strength $A_1$. Each pair of ($A_1,b$) values is chosen such that density $\rho_{\rm bulk}$ in the middle of a 
	film  deviates from the experimental one by  $\sim$0.1$\%$. Each pair ($A_1,d$) or ($A_1,\gamma$) corresponds to the respective ($A_1,b$) pair. Dashed lines connecting the data 
	points are only guides to the eye.}
      	\label{fig:a1vsbS}
\end{figure}

\begin{figure}[h]
	\centering
	\includegraphics[width=100mm]{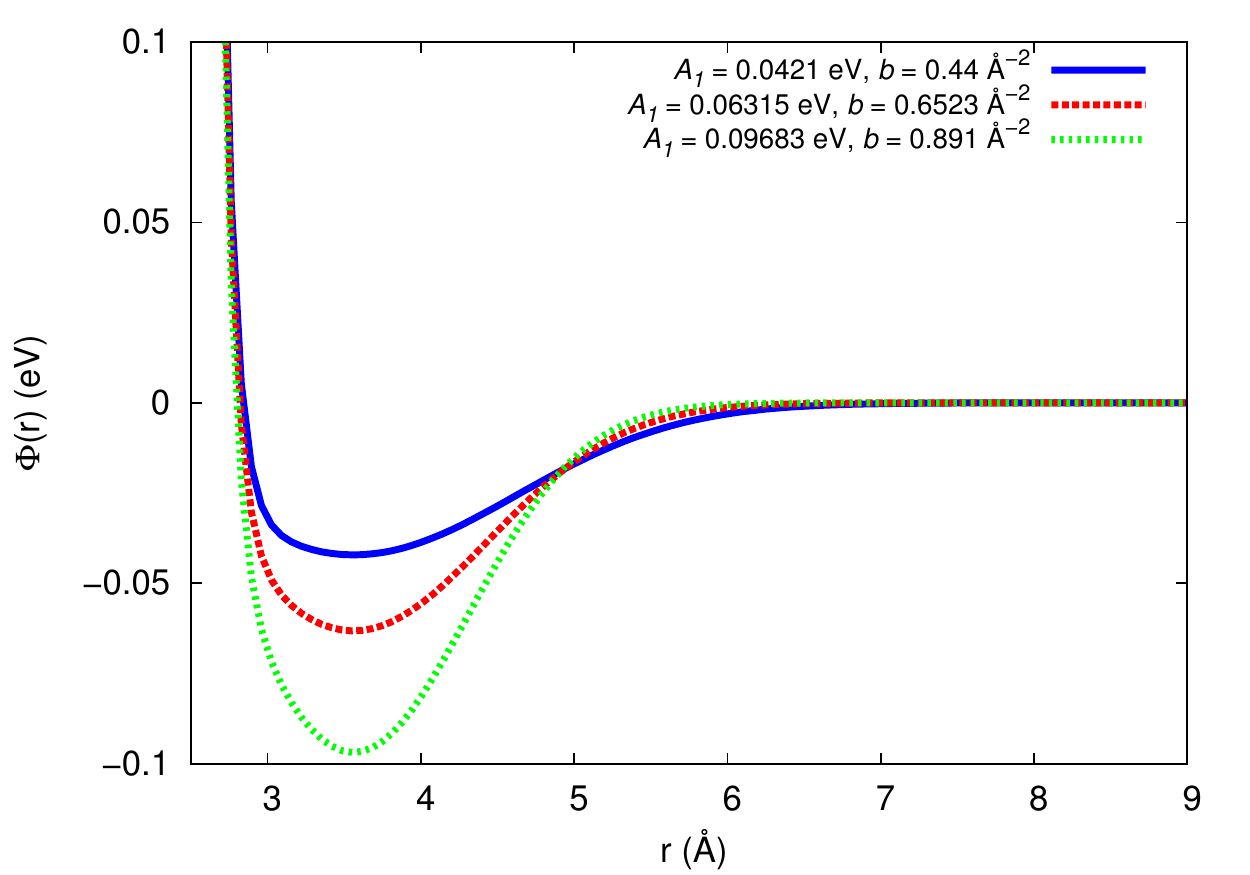}
	\caption{Comparison of our optimized DI (ODI) force field to the DI one. The blue solid line depicts the DI model and the other two  lines (red dashed and green short dashed lines) 
	         show the optimized potential for various parameters as indicated in the key.}
      	\label{fig:optimized}
\end{figure}

The more advanced (compared to LJ) DI model yields $\gamma = 0.18$~N/m at $T = 293$~K, which is larger than the LJ prediction but still smaller in comparison to the experiment. 
In the view of the computational efficiency of the DI model (it is $\sim$2.6 times faster than the DD one), it is interesting to explore the capability of potentilas with the functional form given by 
Eq.~\ref{eq:DI} to yield  a higher surface tension. In the following we demonstrate that it is possible to optimize the DI model to increase the surface tension $\gamma$ while 
preserving the right value of the coexistence density  $\rho_{\rm{coex}}$ of liquid mercury. One can achieve higher $\gamma$ values by increasing the potential depth $A_1$ 
(see the inset in Fig.~\ref{fig:a1vsbS}), but simultaneously one has to adjust the $b$ parameter to preserve the coexistence density of the liquid phase. 
By systematically doing so one decreases the range of the attractive part of the pair interaction as shown in Figs.~\ref{fig:a1vsbS}~and~\ref{fig:optimized}.
As is discussed by Frenkel \textit{et al.} \cite{frenkel}, the range of the attractive interaction determines the stability of liquid phase. A shorter range of attractive interactions tends 
to shift the liquid-crystal phase boundary towards the lower density values for a given temperature. This means there is a limit on how much one can keep decreasing the range of 
attraction before crystallization.

Fig.~\ref{fig:a1vsbS} demonstrates that by increasing $A_1$ and $b$ we increase the hard core diameter $d$ (see Eq.~\ref{eq:dhs}) of Hg atoms, which in turn effectively increases the 
packing fraction $\eta$ of the system. Our best values for the optimized DI (ODI) parameters are $A_1 = 0.09683$~eV and $b = 0.891$~\r{A}$^{-2}$, which corresponds to $\eta=0.643$. 
This value of $\eta$ is just slightly below the value, $\eta=0.65$, for the random closest packing of hard spheres \cite{hansenmc}. If we increase $A_1$ any further at $T = 293$~K, 
we  would inevitably drive the system into the crystalline phase.

Comparing the ODI model to the LJ system, one finds that the packing fraction $\eta$ for both systems in the liquid phase can be higher than that of hard spheres at liquid-solid transition. 
Using Eq.~\ref{eq:dhs} we obtain $d\sigma=1.07$ for the LJ fluid near the triple point. The respective packing fraction $\eta=0.51$, whereas $\eta=0.494$ for the HS fluid at liquid-solid transition \cite{hansenmc}.
\begin{figure}[h]
	\centering
	\includegraphics[width=100mm]{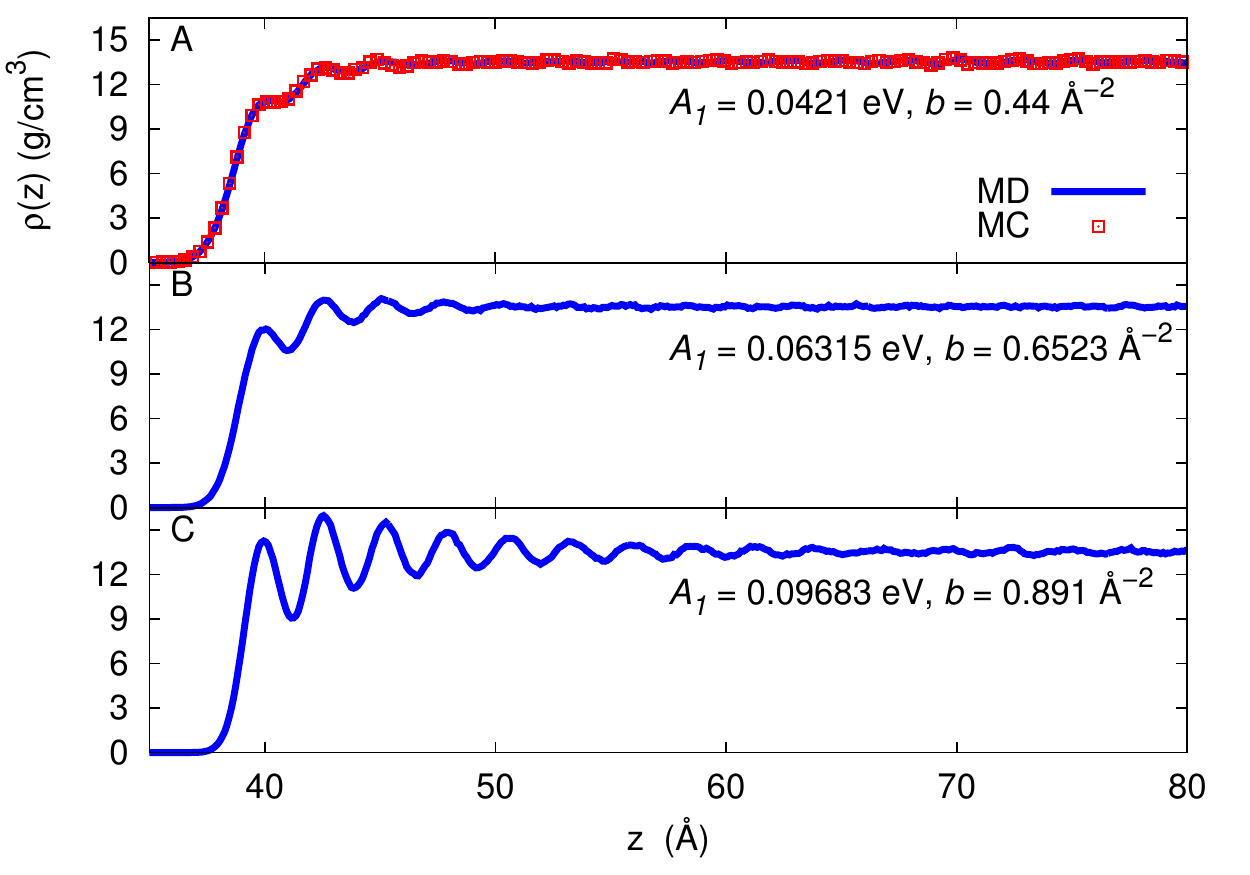}
	\caption{Density profiles for the density-independent model with unmodified (A), modified intermediate (B) and our optimal (C) values of $A_1$ and $b$.}
      	\label{fig:dp_optimized}
\end{figure}
The surface tensions $\gamma$ for the 1st and 2nd sets of the optimized parameters is 0.27 N/m and 0.21 N/m respectively, which are still smaller than the experimental values but, 
nevertheless, larger than the $\gamma$ value for the above tested DI model. In order to further increase $\gamma$ one would have to add additional terms to the pair potential 
(Eq.~\ref{eq:DI}) to keep the range of attraction large enough to prevent the system from crystallizing. 

The density profiles for the DI, intermediate, and the ODI models are shown in Fig.~\ref{fig:dp_optimized}. As it should be the MC density profile overlays perfectly with the MD 
one for the DI model (Fig.~\ref{fig:dp_optimized}A). In all the cases the densities of liquid Hg at the coexistence with vapor are in a very good agreement with the  experimetal value 
of the liquid Hg bulk density of 13.55~g/cm$^3$ at $T = 293$~K \cite{tamura}, and deviate from it on about 0.1$\%$.
The ODI model reveals a much stonger surface layering at the liquid-vapor interface compared to the DI model (Fig.~\ref{fig:dp_optimized}A) confirming that the higher $\gamma$ 
values are related to the stronger layering at the mercury surface. The first outer peak in the density profile for the ODI model (Fig.~\ref{fig:dp_optimized}C) is smaller than the 
second one, whereas in the experiment the opposite behavior is observed \cite{magnussen}. However, this effect should not be significant for the construction of coarse-grained models 
starting from our atomistic ODI force field, because the large scale properties (e.g. contact angle of a drop) are chiefly dictated by the interfacial tension.

\FloatBarrier

\subsection{Density-Dependent Models}
\begin{figure}[h]
	\centering
	\includegraphics[width=85mm]{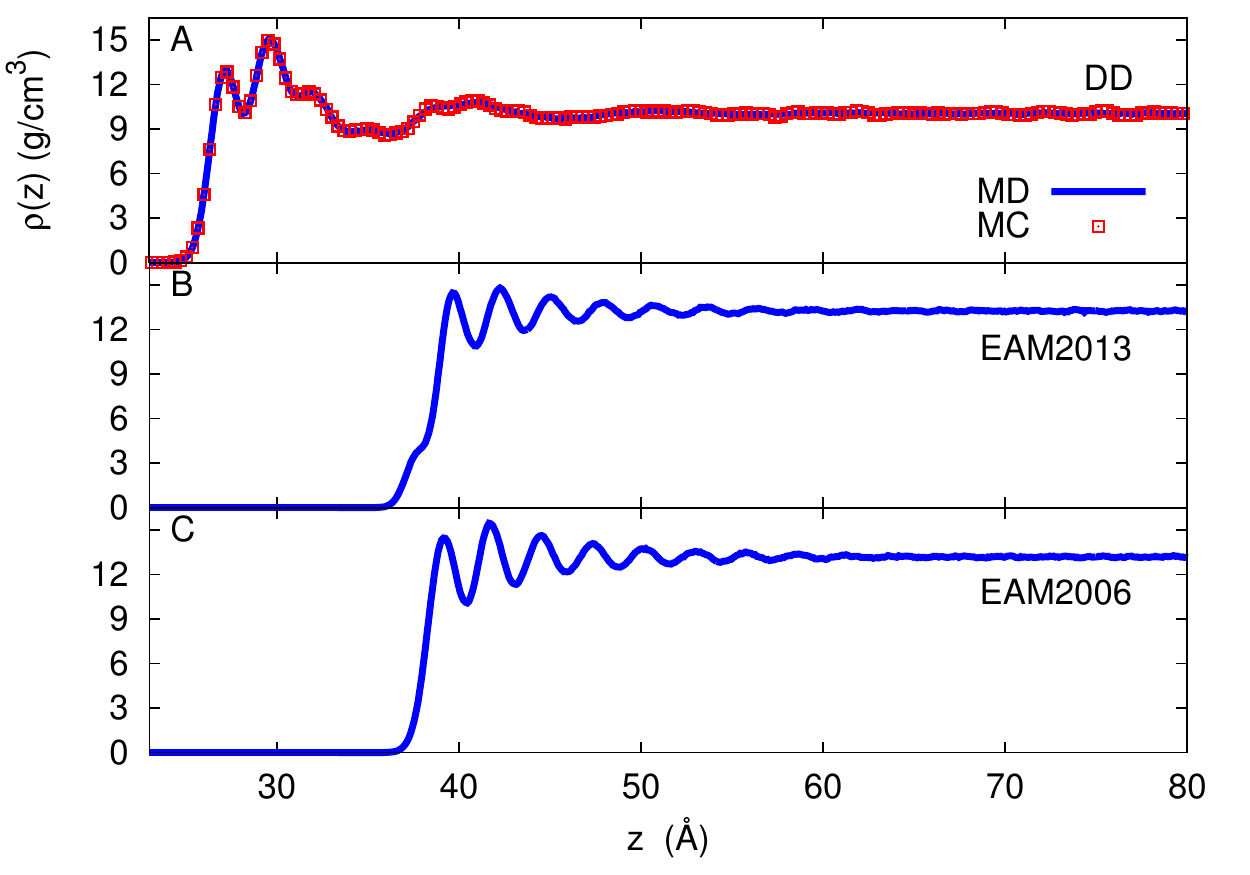}
	\caption{Density profiles for the DD (A), EAM2013 (B) and EAM2006 (C) models of liquid mercury at $T = 293$~K.}
      	\label{fig:dp_multi_eam}
\end{figure}

For the DD, EAM2013 and EAM2006 models we obtain the surface tension $\gamma$ of 0.23, 0.306 and 0.31 N/m, respectively. The EAM models show quite an improvement for the surface tension 
of liquid Hg in comparison to the other models discussed herein. The corresponding density profiles (Fig.~\ref{fig:dp_multi_eam}) also exhibit strong surface layering. 
The MC density profile overlays perfectly with the MD one for the DD model (Fig.~\ref{fig:dp_multi_eam}A).
The obtained Hg liquid densities $\rho_{\rm{coex}}$ in the middle of the Hg film comprise 10.06, 13.25 and 13.18 g/cm$^3$ for the DD, EAM2013 and EAM2006 models,
respectively. The $\rho_{\rm{coex}}$ for the DD is considerably lower compared to the experimental one. 
To exclude possible finite size effects we have also carried out the simulation
of the DD model with $10^4$ atoms and the same area of the Hg film. The result was essencially the same.
In contrast, the EAM models are in a good agreement with the experiment just slightly underestimating $\rho_{\rm{coex}}$. 

\begin{figure}[h]
	\centering
	\includegraphics[width=100mm]{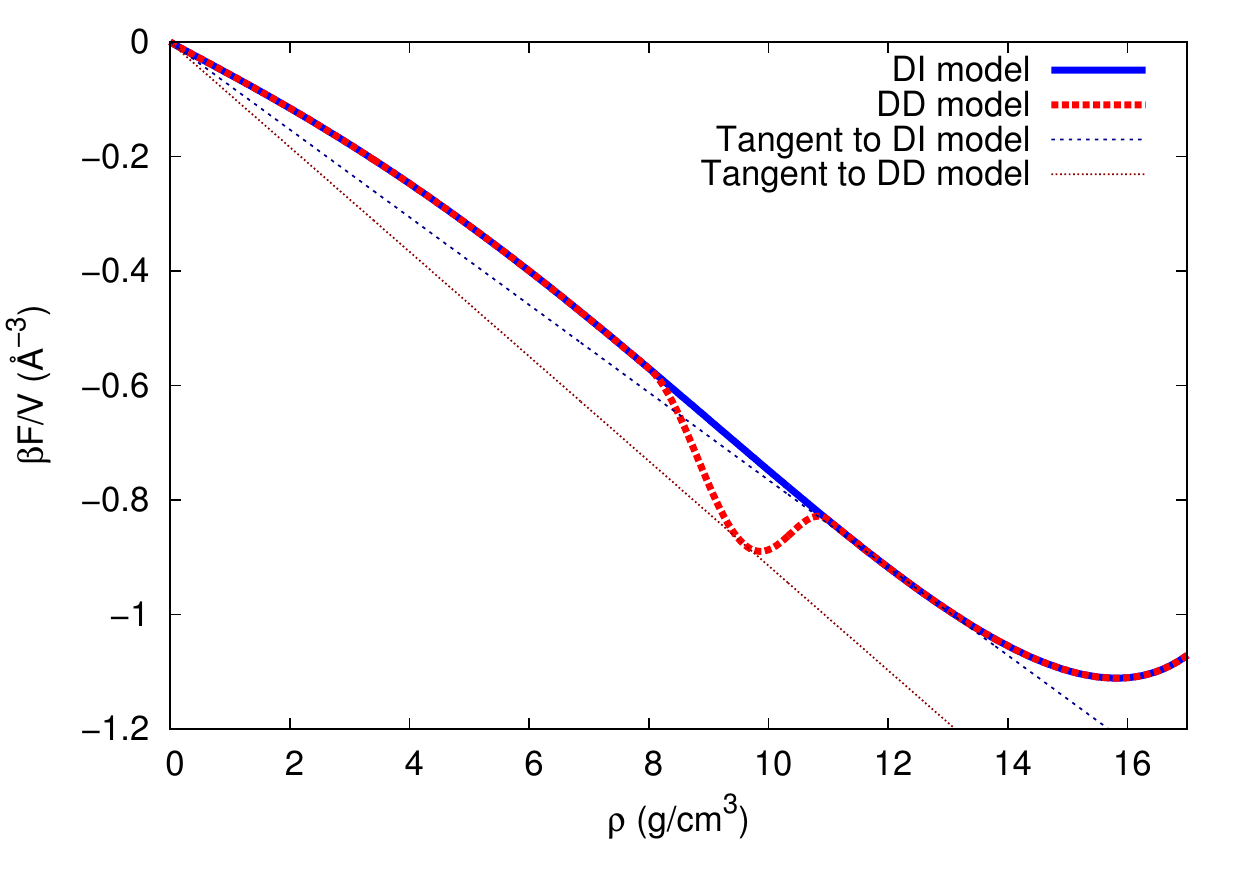}
	\caption{Free energy for the DI (blue solid line) and DD (red dashed line) models. Double tangent to the free energy; DI model: dark blue short dashed line; DD model:
	dark red double short dashed line.}
      	\label{fig:free}
\end{figure}

To rationalize our simulation results we resort to the semi-analytic Liquid State Theory (LST) approach \cite{barker, landau, chang, carnahan, WCA_science, weeks}, described in 
Sec.~\ref{sec:LST}. Let us first consider the Helmholtz free energy (Eq.~\ref{eq:free_en}) for the DD model, which is depicted in Fig.~\ref{fig:free}. 
The free energy for the DI model is also plotted in order to illustrate the effect of the empirical insertion of the density dependence into the DI model (Eq.~\ref{eq:DI}).
The most important difference between the DI and DD models is that the DI model exhibits a single region of negative curvature of $F(\rho)$, which corresponds to the spinodal region 
inside the liquid-vapor miscibility gap. The DD model, however, features two separate regions of negative curvatures separated by a (meta)stable region in the density range 
from 8~g/cm$^3$ to 11~g/cm$^3$. This is exactly the range of densities where the DI pair potential (Eq.~\ref{eq:DI}) is made density-dependent in the empirical DD model 
(see Eq.~\ref{eq:DD}). As we shall see this artifact has dramatic consequences for the overall phase behavior. 
The densities of two thermodynamic states that coexist are determined by the double tangent construction to the free energy (see e.g. Ref.~\cite{cahn}). 
This requirement ensures the equality of the chemical potential and pressure in the coexisting phases. Two systems, with densities equal to those at the points where the double tangent 
touches the free energy curve, will thus be in thermodynamic coexistence. As illustrated in Fig.~\ref{fig:free}, our simple LST predictions yield that the coexistence density 
$\rho_{\rm{coex}}$ of liquid mercury with its vapor is 12.32~g/cm$^3$ and 9.56~g/cm$^3$ for the DI and DD models, respectively. 
These LST results are only qualitative but they clearly demonstrate what happens with the phase behavior when one inserts a density dependence in the pair potential  in an ad-hoc way
and thus rationalize our simulation results. In the case of the DD model the additional dip in the free energy, $F(\rho)$, with the minimum in the vicinity of 10~g/cm$^3$ prevents 
the liquid mercury at a higher density from the coexisting with its vapor, instead there are two coexistence regions. The vapor coexists with a low-density liquid of about 10~g/cm$^3$, 
and there is an additional, spurious liquid-liquid phase coexistence between liquids of density of about 10~g/cm$^3$ and 15~g/cm$^3$. In the simulations, the latter miscibility gap may be
preempted by the coexistence of liquid mercury and its crystalline phase. 
Different definitions of the local density, e.g., via the Gibbs diving surface, may result in a different surface packing and thermodynamics \cite{bomont3}. 

\begin{figure}[h]
	\centering
	\includegraphics[width=85mm]{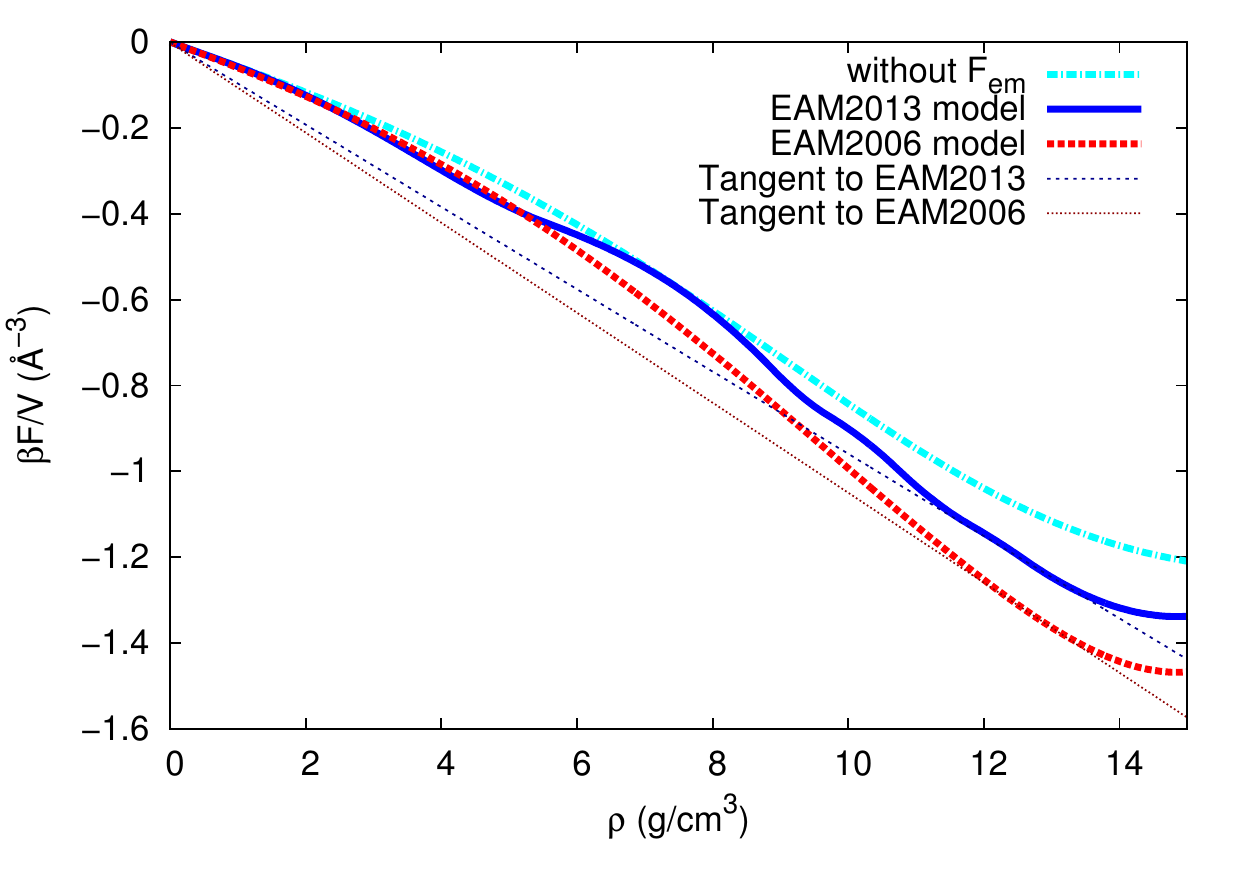}
	\caption{Free energy for the Embedded-Atom Models of mercury.}
      	\label{fig:free_en_eam}
\end{figure}

\begin{figure}[h]
	\centering
	\includegraphics[width=85mm]{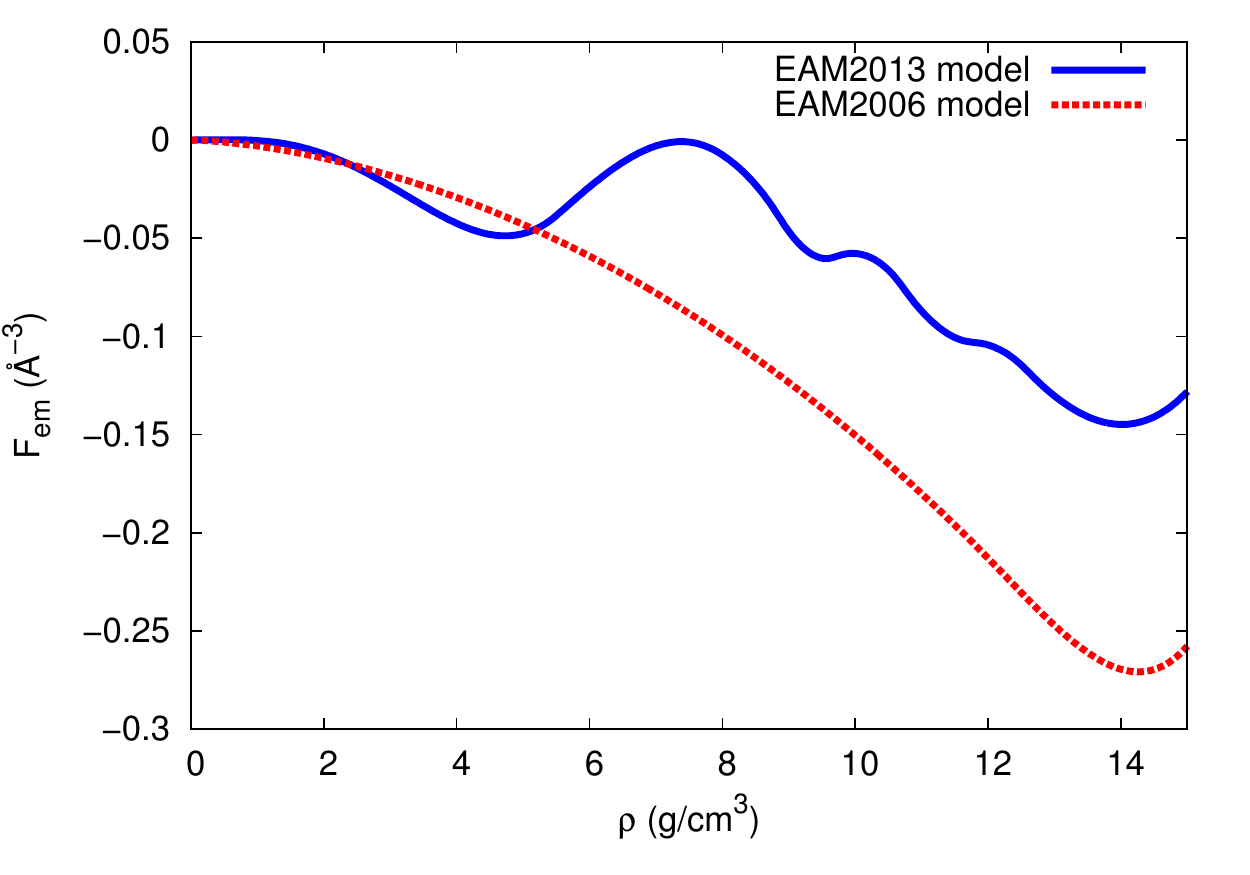}
	\caption{Embedded energy contribution to the free energy.}
      	\label{fig:free_en_eam_density}
\end{figure}

On the contrary, as is shown in Fig.~\ref{fig:free_en_eam} the LST free energy curves for the EAM2013 and EAM2006 models reveal no spurious features and thus yield essentially 
the correct qualitative phase behavior. The free energy for the EAM2013 shows though multiple regions of positive curvature for the value of $\rho$ smaller than 13~g/cm$^3$ indicating 
a number of possible metastable coexisting states at low densities (Fig.~\ref{fig:free_en_eam}). 
Such a behavior of the total free energy might be traced to the corresponding curvature of the embedding free energy $F_{\rm{em}}$ (Fig.~\ref{fig:free_en_eam_density}), 
which in turn stems from the form of the embedding energy $\Phi_{\rm{em}}$ of the EAM2013 model compared to the EAM2006 one (see Fig.~\ref{fig:phi_of_rho_eam}). 
Nevertheless, as we see from the MD simulations, which are confirmed by the LST arguments, all the possible coexistences at lower densities are preempted 
by the liquid-vapor coexistence at the liquid Hg density of 13 g/cm$^3$, and
such a peculiar form of $\Phi_{\rm{em}}$ of the EAM2013 model has basically no qualitative effect neither on the correct thermodynamics of coexistence nor on the surface tension.
The coexistence density thus obtained for the EAM2006 model from the double tangent construction is accordingly 12.8 g/cm$^3$.
For the comparison the free energy without the embedded energy contributions is also depicted in Fig.~\ref{fig:free_en_eam}.
The agreement with the results of the atomistic MD simulations is remarkably good, given the simplicity of the LST approximations. 
The effective hard sphere diameter for both EAM models is independent of density and comprises 2.96~\r{A}.
To increase the reliability of the predictions from the embedded-atom models for liquid Hg one could attempt to fit the embedding energy to the recent experimental 
results for the equation of state of liquid Hg \cite{ayrinhac}. One could also try to adopt a modified embedded-atom method (MEAM) \cite{lee} to the liquid mercury, 
where the non-uniform angle distribution of density is taken explicitly into account. 
Alternatively, one could adopt a systematically modified embedded-atom method (SMEAM) \cite{mueser}, which implicitly accounts for the angle dependence.

\FloatBarrier

\section{Conclusions}
In the present article we have assessed the ability of various state dependent (either through temperature or density) interaction models to predict the surface tension of liquid mercury at $T = 293$~K.
Since density-dependent models generally require more computational efforts we have optimized a density-independent force field in order to increase the value of the surface tension while preserving the experimental coexistence density.
The optimization procedure is outlined in detail. We came to a conclusion that all the models treated in the current work are capable of reproducing the essential unique properties of the surface of liquid Hg, namely
a densely packed liquid phase in coexistence to vapor and a high surface tension (compared to Lennard-Jones systems), which is responsible for the seamless surface and exceptional non-wetting behavior of liquid Hg droplets at $T = 293$~K.
Our study, however, also indicates that it remains a challenge to devise density-independent models that simultaneously yield the experimental values of coexistence density and surface tension.
In particular cases one has to treat with additional care the empirical density dependence of interactions. One can use the Liquid State Theory as a convenient tool to qualitatively explore the effects due to the density dependence prior
to starting a computationally costly atomistic or coarse-grained computer simulation.

\section*{Acknowledgments}
We thank B. Pokroy for stimulating discussions. A.I. and M.M. acknowledge the financial support from the Volkswagen Foundation within the joint German-Israeli program under the grant VW-ZN2726, and D.B. acknowledges the financial 
support from the Alexander von Humboldt Foundation through an Experienced Researcher Fellowship. A.I. also thanks to F. Leonforte and S. Plimpton for useful advice on LAMMPS.


\begin{thebibliography}{999}

\bibitem{saarnio} Saarnio, K.; Frey, A.; Niemi, J. V.; Timonen, H.; R\"{o}nkk\"{o}, T.; Karjalainen, P.; Vestenius, M.; Teinil\"{a}, K.; Pirjola, L.; Niemel\"{a}, V.; Keskinen, J.; H\"{a}yrinen A.; Hillamo, R.
Chemical Composition and Size of Particles in Emissions of a Coal-Fired Power Plant with Flue Gas Desulfurization.
\textit{J. Aerosol Sci.} \textbf{2014}, 73, 14-26.

\bibitem{toxic} Eisler, R. \textit{Mercury Hazards to Living Organisms}; CRC Press: Boca Raton, 2006.

\bibitem{lamp} Kulshreshtha, D. C. \textit{Basic Electrical Engineering}; Tata McGraw-Hill: New Delhi, 2009.

\bibitem{selin} Selin, N. E. Mercury Rising: Is Global Action Needed to Protect Human Health and the Environment? \textit{Environ.: Sci. Policy Sustainable Dev.} \textbf{2005}, 47, 22-35.

\bibitem{pokroy} Pokroy, B.; Aichmayer, B.; Schenk, A. S.; Haimov, B.; Kang. S. H.; Fratzl, P.; Aizenberg, J. Sonication-Assisted Synthesis of Large, High-Quality Mercury Thiolate Single Crystals Directly from Liquid Mercury.
\textit{J. Am. Chem. Soc.} \textbf{2010}, 132, 14355-14357.

\bibitem{tan} Tan, Z.; Sun, L.; Xiang, J.; Zeng, H.; Liu, Z.; Hu, S.; Qiu, J. Gas-Phase Elemental Mercury Removal by Novel Carbon-Based Sorbents. \textit{Carbon} \textbf{2012}, 50, 362-371.

\bibitem{tao} Tao, S.; Li, C.; Fan, X.; Zeng, G.; Lu, P.; Zhang, X.; Wen, Q.; Zhao, W.; Luo, D.; Fan., C. Activated Coke Impregnated with Cerium Chloride Used for Elemental Mercury Removal from Simulated Flue Gas. \textit{Chem. Eng. J.},
\textbf{2012}, 210, 547-556.

\bibitem{disposal} Chen, H.-R.; Chen, C.-C.; Satyanarayana Reddy, A.; Chen, C.-Y.; Li, W. R.; Tseng, M.-J.; Liu, H.-T.; Pan, W.; Maity J. P.; Atla, S. B. Removal of Mercury by Foam Fractionation Using Surfactin, a Biosurfactant.
\textit{Int. J. Mol. Sci.} \textbf{2011}, 12, 8245-8258.

\bibitem{zhao} Zhao, Y.-P.; Wang, Y. Fundamentals and Applications of Electrowetting: A Critical Review. \textit{Rev. Adhesion Adhesives} \textbf{2013}, 1, 114-174.

\bibitem{kutana1} Kutana, A.; Giapis, K. P. Contact Angles, Ordering, and Solidification of Liquid Mercury in Carbon Nanotube Cavities. \textit{Phys. Rev. B} \textbf{2007}, 76, 195444. 

\bibitem{kutana2} Kutana, A.; Giapis, K. P. Atomistic Simulations of Electrowetting in Carbon Nanotubes. \textit{Nano Lett.} \textbf{2006}, 6, 656-661.

\bibitem{kutana3} Chen, J. Y.; Kutana, A.; Collier, C. P.; Giapis, K. P. Electrowetting in Carbon Nanotubes. \textit{Science} \textbf{2005}, 310, 1480-1483.

\bibitem{babayco} Babayco, C. B.; Chang, P. J.; Land, D. P.; Kiehl, R. A.; Parikh, A. N. Evolution of Conformational Order during Self-Assembly of n-Alkanethiols on Hg Droplets: An Infrared Spectromicroscopy Study. \textit{Langmuir}
\textbf{2013}, 29, 8203-8207.

\bibitem{love} Love, J. C.; Estroff, L. A.; Kriebel, J. K.; Nuzzo, R. G.; Whitesides, G. M. Self-Assembled Monolayers of Thiolates on Metals as a Form of Nanotechnology. \textit{Chem. Rev.} \textbf{2005}, 105, 1103-1169.

\bibitem{demoz} Demoz, A.; Harrison, D. J. Characterization and Extremely Low Defect Density Hexadecanethiol Monolayers on Mercury Surfaces. \textit{Langmuir} \textbf{1993}, 9, 1046-1050.

\bibitem{kraack1} Kraack, H.; Tamam, L.; Sloutskin E.; Deutsch, M.; Ocko, B. M. Alkyl-thiol Langmuir Films on the Surface of Liquid Mercury. \textit{Langmuir} \textbf{2007}, 23, 7571-7582.
\bibitem{ocko1} Ocko, B. M.; Kraack, H.; Pershan, P. S.; Sloutskin, E.; Tamam, L.; Deutsch, M. Crystalline Phases of Alkyl-Thiol Monolayers on Liquid Mercury. \textit{Phys. Rev. Lett.} \textbf{2005}, 017802.
\bibitem{deutsch1} Deutsch, M.; Magnussen, O. M.; Ocko, B. M.; Regan, M. J. The Structure of Alkanethiol Films on Liquid Mercury:  an X-Ray Study, \textit{Thin Films} \textbf{1998}, 24, 179-203.
\bibitem{magnussen1} Magnussen, O. M.; Ocko, B. M.; Deutsch, M.; Regan, M. J.; Pershan, P. S.; Abernathy, D.; Gr\"{u}bel G.; Legrand, J.-F. Self-Assembly of Organic Films on a Liquid Metal. \textit{Nature} \textbf{1996}, 384, 250-252.
\bibitem{kraack2} Kraack, H.; Ocko, B. M.; Pershan, P. S.; Sloutskin, E.; Tamam, L.; Deutsch, M. The Structure and Phase Diagram of Langmuir Films of Alcohols on Mercury. \textit{Langmuir} \textbf{2004}, 20, 5386-5395.
\bibitem{kraack3} Kraack, H.; Deutsch, M.; Ocko B. M.; Pershan, P. S. The Structure of Organic Langmuir Films on Liquid Metal Surfaces. \textit{Nucl. Instrum. Meth. Phys. Res. B} \textbf{2003}, 363-370.
\bibitem{kraack4} Kraack, H.; Ocko, B. M.; Pershan, P. S.; Sloutskin, E.; Deutsch, M. Structure of a Langmuir Film on a Liquid Metal Surface. \textit{Science} \textbf{2002}, 298, 1404-1407.
\bibitem{kraack5} Kraack, H.; Ocko, B. M.; Pershan, P. S.; Tamam, L.; Deutsch, M. Temperature Dependence of the Structure of Langmuir Films of Normal-Alkanes on Liquid Mercury. \textit{J. Chem. Phys.} \textbf{2004}, 121, 8003-8009.
\bibitem{kraack6} Kraack, H.; Ocko, B. M.; Pershan, P. S.; Sloutskin, E.; Deutsch, M. Langmuir Films of Normal-Alkanes on the Surface of Liquid Mercury. \textit{J. Chem. Phys.} \textbf{2003}, 119, 10339-10349.
\bibitem{kraack7} Kraack, H.; Ocko, B. M.; Pershan, P. S.; Sloutskin, E.; Tamam, L.; Deutsch. M. Fatty Acid Langmuir Films on Liquid Mercury:  X-Ray and Surface Tension Studies. \textit{Langmuir} \textbf{2004}, 20, 5375-5385.

\bibitem{stevenson} Stevenson, K. J.; Mitchell, M.; White, H. S. Oxidative Adsorption of n-Alkanethiolates at Mercury. Dependence of Adsorption Free Energy on Chain Length. \textit{J. Phys. Chem. B} \textbf{1998}, 102, 1235-1240.

\bibitem{holmlin} Holmlin, R. E.; Haag, R.; Chabinyc, M. L.; Ismagilov, R. F.; Cohen, A. E.; Terfort, A.; Rampi, M. A.; Whitesides, G. M. Electron Transport through Thin Organic Films in Metal-Insulator-Metal Junctions Based on 
Self-Assembled Monolayers. \textit{J. Am. Chem. Soc.} \textbf{2001}, 123, 5075-5085.

\bibitem{zhu} Zhu, L.; Popoff, R. T. W.; Yu, H.-Z. Metastable Molecular Metal-Semiconductor Junctions. \textit{J. Phys. Chem. C} \textbf{2015}, 119, 1826-1831.

\bibitem{popoff} Popoff, R. T.; Kavanagh, K. L.; Yu, H.-Z. Preparation of Ideal Junctions: Depositing Non-invasive Gold Contacts on Molecularly Modified Silicon. \textit{Nanoscale}, \textbf{2011}, 3, 1434-1445.

\bibitem{weiss} Weiss, E. A.; Kriebel, J. K.; Rampi, M.-A.; Whitesides, G. M. The Study of Charge Transport through Organic Thin Films: Mechanism, Tools and Applications. \textit{Philos. Trans. R. Soc. A} \textbf{2007}, 365, 1509-1537.

\bibitem{rampi} Rampi, M. A.; Schuller, O. J. A.; Whitesides G. M. Alkanethiol Self-Assembled Monolayers as the Dialectric of Capacitors with Nanoscale Thickness. \textit{Appl. Phys. Lett} \textbf{1998}, 72, 1781-1783.

\bibitem{nitzan} Nitzan, A.; Ratner, M. A. Electron Transport in Molecular Wire Junctions. \textit{Science} \textbf{2003}, 300, 1384-1389.

\bibitem{slowinski1} Slowinski, K.; Fong, H. K. Y.; Majda, M. Mercury-Mercury Tunneling Junctions. 1. Electron Tunneling Across Symmetric and Asymetric Alkanethiolate Bilayers. \textit{J. Am. Chem. Soc.} \textbf{1999}, 121, 7257-7261.

\bibitem{slowinski2} Slowinski, K.; Chamberlain, R. V.; Miller, C. J.; Majda, M. Through-Bond and Chain-to-Chain Coupling. Two Pathways in Electron Tunneling through Liquid Alkanethiol Monolayers on Mercury Electrodes. 
\textit{J. Am. Chem. Soc.} \textbf{1997}, 119, 11910-11919.


\bibitem{kiehl} Kiehl, R. A.; Le, J. D.; Candra, P.; Hoye R. C.; Hoye, T. R. Charge Storage Model for Hysteretic Negative-Differential Resistace in Metal-Molecule-Metal Junctions. \textit{Appl. Phys. Lett.} \textbf{2006}, 88, 172102-172104

\bibitem{clarkson} Clarkson, T. W. Human Toxicology of Mercury. \textit{J. Trace Elem. Exp. Med.} \textbf{1998}, 11, 303-317.

\bibitem{zahir} Zahir, F.; Rizwi, S. J.; Haq, S. K.; Khan, R. H. Low Dose Mercury Toxicity and Human Health. \textit{Environ. Toxicol. Phar.} \textbf{2005}, 20, 351-360.

\bibitem{filippini} Filippini, G.; Bonal, C.; Malfrey, P. Atomistic and Energetic Description of Self-Assembled Monolayers of Differently Endgroup-Functionalized Alkanethiols Adsorbed on the Gold Substrate by Using Molecular Simulations.
\textit{Soft Matter} \textbf{2013}, 9, 5099-5109.
\bibitem{ahn} Ahn, Y.; Saha, J. K.; Schatz, G. C.; Jang, J. Molecular Dynamics Study of the Formation of a Self-Assembled Monolayer on Gold. \textit{J. Phys. Chem. C} \textbf{2011}, 115, 10668-10674.
\bibitem{jimenez} Jim\'{e}nez, A.; Sarsa, A.; Bl\'{a}zquez, M.; Pineda, T. A Molecular Dynamics Study of the Surfactant Surface Density of Alkanethiol Self-Assembled Monolayers on Gold Nanoparticles as a Function of the Radius.
\textit{J. Phys. Chem. C} \textbf{2010}, 114, 21309-21314.
\bibitem{ghorai} Ghorai, P. K.; Glotzer, S. C. Molecular Dynamics Simulation Study of Self-Assembled Monolayers of Alkanethiol Surfactants on Spherical Gold Nanoparticles. \textit{J. Phys. Chem. C} \textbf{2007}, 11, 15857-15862.
\bibitem{singh} Singh, C.; Ghorai, P. K.; Horsch, M. A.; Jackson, A. M.; Larson, R. G. Entropy-Mediated Patterning of Surfactant-Coated Nanoparticles and Surfaces. \textit{Phys. Rev. Lett.} \textbf{2007}, 99, 226106.
\bibitem{henz} Henz, B. J.; Zachairah, M. R. Molecular Dynamics Study of Alkanethiolate Self-Assemble Monolayer Coated Gold Nanoparticle. \textit{HPCMP-UGC} \textbf{2007}, 0-7695-3088-5/07.
\bibitem{rai} Rai, B.; Chetan, S. P.; Malhotra, C. P.; Ayappa, K. G. Molecular Dynamics Simulations of Self-Assembled Alkylthiolate Monolayers on an Au(111) Surface. \textit{Langmuir} \textbf{2004}, 20, 3138-3144.
\bibitem{vemparala} Vemparala, S.; Karki B. B. Large-Scale Molecular Dynamics Simulations of Alkanethiol Self-Assembled Monolayers. \textit{J. Chem. Phys.} \textbf{2004}, 121, 4323-4330.
\bibitem{zhang} Zhang, L.; Goddard III, W. A.; Jiang, S. Molecular Simulation Study of the c(4$\times$2) Superlattice Structure of Alkanethiol Self-Assembled Monolayers on Au(111). \textit{J. Chem. Phys.} \textbf{2002}, 117, 7342-7349.
\bibitem{luedtke} Luedtke, W. D.; Landman, U. Structure and Thermodynamics of Self-Assembled Monolayers on Gold Nanocrystallites. \textit{J. Phys. Chem. B} \textbf{1998}, 102, 6566-6572.
\bibitem{bhatia1} Bhatia, R.; Garrison, B. J. Structure of c(4$\times$2) Superlattice in Alkanethiolate Self-Assemble Monolayer. \textit{Langmuir} \textbf{1997}, 13, 4038-4043.
\bibitem{bhatia2} Bhatia, R.; Garrison, B. J. Phase Transition in Methyl-Terminated Monolayer Self-Assembled on Au\{111\}. \textit{Langmuir} \textbf{1997}, 13, 765-769.
\bibitem{tupper} Tupper, K. J.; Colton, R. J.; Brenner, D. W. Simulation of Self-Assembled Monolayers under Compression: Effect of Surface Asperities. \textit{Langmuir} \textbf{1994}, 10, 2041-2043.
\bibitem{sellers} Sellers, H.; Ulman, A.; Shnidman, Y.; Eilers, J. E. Structure and Binding of Alkathiolates on Gold and Silver Surfaces: Implications for Self-Assembled Monolayers. \textit{J. Am. Chem. Soc.} \textbf{1993}, 115, 9389-9401.
\bibitem{hautman} Hautman, J; Klein, M. L. Simulation of a Monolayer of Alkyl Thiol Chains. \textit{J. Chem. Phys.} \textbf{1989}, 91, 4994-5001.

\bibitem{desgranges} Desgranges, C.; Delhommelle, J. Thermodynamics of Phase Coexistence and Metal-Nonmetal Transition in Mercury: Assessment of Effective Potential via Expanded Wang-Landau Simulations. \textit{J. Phys. Chem. B} \textbf{2014}, 118, 3175-3182.
\bibitem{belashch1} Belashchenko, D. K. Application of the Embedded Atom Model to Liquid Mercury. \textit{High Temperature} \textbf{2013}, 51, 40-48.
\bibitem{chacon} Chac\'{o}n, E.; Reinaldo-Falag\'{a}n, M.; Velasco, E.; Tarazona, P. Layering of Free Liquid Surface. \textit{Phys. Rev. Lett.} \textbf{2001}, 166101.
\bibitem{bomont1} Bomont, J.-M.; Bretonnet, J.-L.; Gonzalez, D. J.; L. E. Gonzalez. Computer Simulation Calculations of the Free Liquid Surface of Mercury. \textit{Phys. Rev. B} \textbf{2009}, 79, 144202.
\bibitem{belashch2} Belashchenko, D. K. Application of the Embedded Atom Model to Liquid Metals: Liquid Mercury. \textit{High Temperature} \textbf{2006}, 44, 675-686.
\bibitem{raabe1} Raabe, G.; Todd, B. D.; Sadus, R. J. Molecular Simulation of the Shear Viscosity and Self-Diffusion Coefficient of Mercury along the Vapor-Liquid Coexistence Curve. \textit{J. Chem. Phys.} \textbf{2005}, 123, 034511.
\bibitem{ghatee} Ghatee, M. H.; Bahadori, M. Density-Dependent Equations of State for Metal, Nonmetal, and Transition States for Compressed Mercury Fluid. \textit{J. Phys. Chem. B} \textbf{2004}, 108, 4141-4146.
\bibitem{raabe2} Raabe, G.; Sadus, R. J. Molecular Simulation of the Vapor-Liquid Coexistence of Mercury. \textit{J. Chem. Phys.} \textbf{2003}, 119, 6691-6697.
\bibitem{toth} T\'{o}th, G. An Iterative Scheme to Derive Pair Potentials from Structure Factors and its Application to Liquid Mercury. \textit{J. Chem. Phys.} \textbf{2003}, 118, 3949-3955.
\bibitem{okumura} Okumura, H.; Yonezawa, F. Bulk Viscosity in a Density-Dependent-Potential System. \textit{J. Non-Cryst. Sol.} \textbf{2002}, 260-264.
\bibitem{belashch3} Belashchenko, D. K. The Simulation of Liquid Mercury by Diffracation Data and the Inference of the Interparticle Potential. \textit{High Temperature} \textbf{2002}, 40, 212-221.
\bibitem{sumi} Sumi, T.; Miyoshi, E.; Tanaka, K. Molecular-Dynamics Study of Liquid Mercury in the Density Region Between Metal and Nonmetal. \textit{Phys. Rev. B} \textbf{1999}, 59, 6153-6158.
\bibitem{munejiri} Munejiri, S.; Shimojo, F.; Hoshino, K. The Density Dependence of the Velocity of Sound in Expanded Liquid Mercury Studied by Means of a Large-Scale Molecular-Dynamics Simulations. \textit{J. Phys.: Condens. Matter} \textbf{1998}, 10, 4963-4974.

\bibitem{louis} Louis, A. A. Beware of Density Dependent Pair Potentials. \textit{J. Phys.: Condens. Matter} \textbf{2002}, 14, 9187-9206.

\bibitem{shen} Shen, K. S.; Mountain, R. D.; Errington, J. R. Comparative Study of the Effect of Tail Corrections on Surface Tension Determined by Molecular Simulation. \textit{J. Phys. Chem. B} \textbf{2007}, 111, 6198-6207.
\bibitem{johnson} Johnson, J. K.; Zollweg, J. A.; Gubbins, K. E. The Lennard-Jones Equation of State Revisited. \textit{Mol. Phys.} \textbf{1993}, 3, 591-618.
\bibitem{nijmeijer} Nijmeijer, M. J. P.; Bakker, A. F.; Bruin, C.; Sikkenk, J. H. A Molecular Dynamics Simulation of the Lennard-Jones Liquid-Vapor Interface. \textit{J. Chem. Phys.} \textbf{1988}, 89, 3789-3792.
\bibitem{gast} Adamson, A. W.; Gast, A. P. \textit{Physical Chemistry of Surfaces, 6th ed.}; John Wiley \& Sons: New York, 1997.
\bibitem{kutana4} See the suplimental matrial for Ref.~\cite{kutana3}.
\bibitem{hoshino} Hoshino, K.; Tanaka, S.; Shimojo, F. Dynamical Structure of Fluid Mercury: Molecular-Dynamics Simulations.  \textit{J. Non-Cryst. Sol.} \textbf{2007}, 353, 3389-3393.
\bibitem{bomont2} Bomont, J.-M.; Bretonnet, J.-L. An Effective Pair Potential for Thermodynamics and Structural Properties of Liquid Mercury. \textit{J. Chem. Phys.} \textbf{2006}, 124, 054504.

\bibitem{rapaport} Rapaport, D. C. \textit{The Art of Molecular Dynamics Simulation}; Cambridge University Press: Cambridge, 2005.

\bibitem{foiles1} Foiles, S. M. Application of the Embedded-Atom Method to Liquid Transition Metals. \textit{Phys. Rev. B} \textbf{1985}, 32, 3409-3415.
\bibitem{foiles2} Foiles, S. M.; Adams, J. B. Termodynamic Properties of FCC Transion Metals as Calculated with the Embedded-Atom Method. \textit{Phys. Rev. B} \textbf{1989}, 40, 5909-5915.
\bibitem{holzman} Holzman, L. M.; Adams, J. B.; Foiles, S. M.; Hitchon, W. N. G. Properties of the Liquid-Vapor Interface of FCC Metals Calculated Using the Embedded Atom Method. \textit{J. Mater. Res.} \textbf{1991}, 6, 298-302.

\bibitem{lammps} Plimpton, S. Fast Parallel Algorithms for Short-Range Molecular Dynamics. \textit{J. Comp. Phys.} \textbf{1995}, 117, 1-18. \newline
		  \texttt{http://lammps.sandia.gov}

\bibitem{nose} Nos\'{e}, S. A Unified Formulation of the Constant Temperature Molecular Dynamics Methods. \textit{J. Chem. Phys.} \textbf{1984}, 81, 511-519.

\bibitem{hoover} Hoover, W. G. Canonical Dynamics: Equilibrium Phase-Space Distributions. \textit{Phys. Rev. A} \textbf{1985}, 31, 1695-1697.

\bibitem{chains} Martyna, G. J.; Klein, M. L.; Tuckerman, M. Nos\'{e}–Hoover Chains: The Canonical Ensemble via Continuous Dynamics. \textit{J. Chem. Phys.} \textbf{1992}, 97, 2635-2643.

\bibitem{kirkwood} Kirkwood, J. G.; Buff, F. P. The Statistical Mechanical Theory of Surface Tension. \textit{J. Chem. Phys.} \textbf{1949}, 17, 338-343.

\bibitem{tildesley} Allen, M. P.;  Tildesley, D. J. \textit{Computer Simulation of Liquids}; Clarendon Press: Oxford, 1991.

\bibitem{metropolis} Metropolis, N.; Rosenbluth, A. W.; Rosenbluth M. N.; Teller, A. H. Equation of State Calculations by Fast Computing Machines. \textit{J. Chem. Phys.} \textbf{1953}, 21, 1087-1092.

\bibitem{kitamura} Kitamura, H. Equation of State for Expanded Fluid Mercury: Variational Theory with Many-Body Interaction. \textit{J. Chem. Phys.} \textbf{2007}, 126, 134509.
\bibitem{barker} Barker, J. A.; Henderson, D. Pertubation Theory and Equation of State for Fluids. II. A Successful Theory of Liquids. \textit{J. Chem. Phys.} \textbf{1967}, 47, 4714-4720.
\bibitem{chang} Chang. J.; Sandler, S. I. A Real Function Representation for the Structure of the Hard-Spheres Fluid. \textit{Mol. Phys.} \textbf{1994}, 81, 735-744.
\bibitem{landau} Landau, L. D.; Lifshitz, E. M. \textit{Statistical Physics, part 1, 3rd ed.}; BPC Wheatons: Exeter, 1994.
\bibitem{carnahan} Carnahan, N. F.; Starling, K. E. Equation of State for Nonattracting Rigid Spheres. \textit{J. Chem. Phys.} \textbf{1969}, 51, 635-636.
\bibitem{WCA_science} Chandler, D.; Weeks, J. D.; Andersen, H. C. Van der Waals Picture of Liquids, Solids, and Phase Transformations. \textit{Science} \textbf{1983}, 220, 787-794.
\bibitem{weeks} Weeks, J. D.; Chandler, D.; Andersen, H. C. Role of Repulsive Forces in Determining the Equilibrium Structure of Simple Liquids. \textit{J. Chem. Phys.} \textbf{1971}, 54, 5237-5246.
\bibitem{tamura} Tamura, K.; Hosokawa, S. Structure Studies of Expanded Fluid Mercury up to the Liquid-Vapor Critical Region. \textit{Phys. Rev. B} \textbf{1998}, 58, 9030-9038.
\bibitem{cahn} Cahn, J. W.; Hilliard, J. E. Free Energy of a Nonuniform System. I. Interfacial Free Energy. \textit{J. Chem. Phys.} \textbf{1958}, 28, 258-267.

\bibitem{frenkel} Frenkel, D.; Bladon, P.; Bolhuis, P.; Hagen, M. Liquid-Like Behavior in Solids. \textit{Mol. Sim.} \textbf{1996}, 16, 127-137.

\bibitem{hansenmc} Hansen, J.-P.; McDonald, I. R. \textit{Theory of Simple Liquids, 3rd ed.}; Academic Press: London, 2006.

\bibitem{magnussen} Magnussen, O. M.; Ocko, B. M.; Regan, M. J.;  Penanen, K.; Pershan P. S.; Deutsch, M. X-Ray Reflectivity Measurements of Surface Layering in Liquid Mercury. \textit{Phys. Rev. Lett.} \textbf{1995}, 74, 4444-4447.

\bibitem{bomont3} Bomont, J.-M. Private communication.

\bibitem{ayrinhac} Ayrinhac, S.; Bove, L. E.; Morand, M.; Le Marchand, G.; Bergame, F.; Philippe, J.; Decremps, F. Equation of State of Liquid Mercury to 520 K and 7 GPa from Acoustic Velocity Measurements. \textit{J. Chem Phys.} 
\textbf{2014}, 140, 244201.

\bibitem{lee} Lee, B.-J.; Shim, J.-H.; Baskes, M. I. Semiempirical Atomic Potentials for the FCC Metals Cu, Ag, Au, Ni, Pd, Pt, Al, and Pb Based on First and Second Nearest-Neighbor Modified Embedded Atom Method. \textit{Phys. Rev. B}
\textbf{2003}, 68, 144112.

\bibitem{mueser} Jalkanen, J.; M\"{u}ser, M. H. Systematic Analysis and Modification of Embedded-Atom Potentials: Case Study of Copper. \textbf{2014}.
 Pre-print available at \texttt{http://www.lms.uni-saarland.de/wp-content/uploads/2014/11/14EAM\_MSMSE.pdf}
 
\end{thebibliography}
\end{document}